\documentclass[12pt, preprint]{emulateapj}

\newcommand\um{\ifmmode{\mu{\rm m}}\else{$\mu$m}\fi}

\begin{document}
\title{Isotropic Mid-Infrared Emission from the Central 100 pc of Active Galaxies}
\author{
N. A. Levenson\altaffilmark{1,2}, 
J. T. Radomski\altaffilmark{2}, 
C. Packham\altaffilmark{3}, 
R. E. Mason\altaffilmark{4}, 
J. J. Schaefer\altaffilmark{3},  
and C. M. Telesco\altaffilmark{3} 
}
\altaffiltext{1}{Department of Physics and Astronomy, University of Kentucky,
Lexington, KY 40506; nlevenson@gemini.edu}
\altaffiltext{2}{Gemini Observatory, Casilla 603, La Serena, Chile; jradomski@gemini.edu}
\altaffiltext{3}{Department of Astronomy, University of Florida, 
211 Bryant Space Science Center, P.O. Box 112055, Gainesville, FL 32611;
packham@astro.ufl.edu; justin.schaefer@seakr.com; telesco@astro.ufl.edu}
\altaffiltext{4}{Gemini Observatory, 670 N. A'ohoku Place, Hilo, HI 96720; rmason@gemini.edu}
\submitted{ }
\journalinfo{Accepted for publication in the Astrophysical Journal}
\begin{abstract}
  Dust reprocesses the intrinsic radiation of active galactic nuclei
  (AGNs) to emerge at longer wavelengths.  The observed mid-infrared
  (MIR) luminosity depends fundamentally on the luminosity of the
  central engine, but in detail it also depends on the geometric
  distribution of the surrounding dust.  To quantify this
  relationship, we observe nearby normal AGNs in the MIR to achieve
  spatial resolution better than 100 pc, and we use
  absorption-corrected X-ray luminosity as a proxy for the intrinsic
  AGN emission.  We find no significant difference between optically
  classified Seyfert 1 and 2 galaxies.  Spectroscopic differences,
  both at optical and IR wavelengths, indicate that the immediate
  surroundings of AGNs is not spherically symmetric, as in standard
  unified AGN models.  A quantitative analysis of clumpy torus
  radiative transfer models shows that a clumpy local environment can
  account for this dependence on viewing geometry while producing MIR
  continuum emission that remains nearly isotropic, as we observe,
  although the material is not optically thin at these wavelengths.
  We find some luminosity dependence on the X-ray/MIR correlation in
  the smallest scale measurements, which may indicate enhanced dust
  emission associated with star formation, even on these sub-100 pc
  scales.

\end{abstract}

\section{Introduction}

While active galactic nuclei (AGNs) present a variety of observational
characteristics, unified models suggest that the central engines of all these
objects are fundamentally the same.  Accretion onto a supermassive black hole
is the basic energy source, and an optically and geometrically thick surrounding
``torus'' introduces effects that depend on viewing geometry.
In particular, spectrally broad emission lines are observed in type 1 AGNs,
which afford a direct view of the fast-moving material close to the central engine,
while the torus blocks the view of this broad line region in type 2 objects
\citep{Ant93}.

The dust in the obscuring torus also reprocesses the intrinsic hard
continuum radiation to longer wavelengths, and the bulk of it emerges
in the mid-infrared (MIR) regime (5--30\um).  
The observed MIR emission of AGNs
thus is sensitive to the intrinsic bolometric luminosity of the AGN.
High energy X-ray emission reveals the bolometric luminosity of the
AGNs, and few other energy sources contribute substantially to these
observed fluxes.  Provided that measured 2--10 keV luminosities are
corrected for intrinsic absorption, which can be significant, this
X-ray luminosity is an effective proxy for the AGN bolometric
luminosity, representing approximately 5\% of $L_{bol}$ \citep{Elv94}.

In detail, the spectral energy distribution of the reprocessed
emission also depends on the geometric distribution of the dust.   
If the torus material is smoothly distributed, for
example, the observed IR flux along unobscured lines of sight to the hot,
optically-thin surface of the inner torus is much greater than
that measured along lines of sight through the optically thick torus
\citep{Pier92,Gra94,Efs95}.  In contrast, if the dusty material is
separated into clumps that do not fill the torus volume, the
luminosity dependence on viewing angle is reduced \citep{Hon06,Nen08b,Scha08}.
MIR photometry of type 1 and 2 AGNs can therefore
 sensitively discriminate between smooth and clumpy dust distributions.
Moreover, recent observations at high spatial resolution
\citep{Jaf04,Pac05,Tri07,Rad08} suggest that the spatial scale of the dusty
torus is small ($< 10$ pc).  Thus, measurements of the torus emission
must be made over limited physical regions, and comparison models must
accommodate the small torus extent.

Previous work generally indicates that the intrinsic and reprocessed
AGN emission are strongly correlated.  However, some of these studies
suffer from poor angular resolution. \citet{Lutz04} and \citet{Ram07},
for example, use MIR data from ISO.  The spatial resolution of these
observations is typically 2 kpc, and dust emission associated with
star formation can contribute greatly on these scales.  Observations
with 8m class telescopes on the ground offer an order of magnitude
greater angular resolution, but initial investigations provided only
small samples \citep{Kra01,Hor06}.  Extending the samples
\citep[e.g.,][]{Hor08,Gan09} introduced mixed classes of AGNs
(including radio galaxies and LINERs, for example) which may have
distinct properties because of their radio loudness and hardness of
the ionizing continuum.  Despite these complications, no earlier work
shows significant differences between type 1 and 2 AGNs in the MIR.
The spectral energy distributions of type 1 and 2 AGNs are
characteristically different through the near-IR, with the former
emitting much more strongly.  The only suggestion of a difference
between types in the MIR emerges at relatively short wavelengths
\citep[$\lambda=6.7\um$;][]{Ram07} and is not maintained in the
longer-wavelength data of the same study (C. Ramos Almeida, private
communication 2009).

In this paper we present MIR imaging of a sample of 17 Seyfert
galaxies to compare with X-ray measurements of the AGN bolometric
luminosity.  We restrict this study to normal Seyfert galaxies,
excluding radio-loud sources, in which non-thermal synchrotron
emission can contribute significantly to the MIR flux.  We require the
sample galaxies to be closer than 50 Mpc, to achieve spatial
resolution around 100 pc.  In contrast with other work, we do not
consider LINER galaxies, which may have unusual MIR properties
\citep{Eli06,Per01}.  Moreover, we take particular care to correct the
observed X-ray fluxes to recover the intrinsic AGN emission.  The
X-rays advantageously offer observations of emission that originates
close to the black hole, avoiding the uncertainties of extinction
corrections and variations among the more distant narrow line regions
of optical [\ion{O}{3}] measurements, for example.  In addition, we
directly compare our MIR results with radiative transfer calculations
of dust emission.

\section{Observations and Data Reduction}

\subsection{MIR Measurements\label{subsec:mirdata}}

Observations were made using T-ReCS \citep{Tel98} on the 8.1m Gemini
South, Michelle \citep{Gla97} on Gemini North, and OSCIR on the Blanco
4m at CTIO, Gemini North, and Gemini South.  Table \ref{tab:obs} shows
the log of observations.  OSCIR used a Boeing $128\times 128$ pixel
Si:As BIB detector, providing a plate scale of 0.183$\arcsec$ per
pixel on the 4m at CTIO.  The spatial resolution achieved was around
1$\arcsec$. Both T-ReCS and Michelle use a Raytheon $320 \times 240$
pixel Si:As IBC array, providing a plate scale of 0.089 and
0.1005$\arcsec$ per pixel respectively, and achieved a spatial
resolution of between 0.3 to 0.5$\arcsec$.  In the case of T-ReCS, the
detector was used in correlated quadruple sampling (CQS) mode
\citep{Sako03}.  In all cases, images were obtained using the standard
chop-nod technique to remove time-variable sky background, telescope
thermal emission and so-called ``$1/f$'' detector noise.  For the
T-ReCS observations, the chop throw was 15$\arcsec$ and the telescope
was nodded every 30 s, whereas Michelle was nodded every 45 s.  The
OSCIR observations had a chop throw of 30$\arcsec$ on the 4m, but
15$\arcsec$ on Gemini, and all telescopes were nodded every 30 s.  The
chop throw was typically at 0$\degr$ (N-S), but was angled in the case
of extended objects.  Observations were made using the N ($\lambda_c =
10.36\um$, $\Delta\lambda = 5.27\um$), N$^\prime$ ($\lambda_c =
11.2\um$, $\Delta\lambda = 2.4\um$), Si2 ($\lambda_c = 8.74\um$,
$\Delta\lambda = 0.78\um$), or Si5 ($\lambda_c = 11.6\um$,
$\Delta\lambda = 1.1\um$) filters, where $\lambda_c$ is the central
wavelength, and the filter width $\Delta\lambda$ indicates 50\% cut-on
and cut-off values.

Data were reduced using in-house developed IDL or equivalent Gemini
IRAF routines.  The difference for each chopped pair for each given
nod-set was calculated, and the nod-sets were then differenced and
combined to create a single image.  Chopped pairs obviously
compromised by cirrus, high electronic noise, or other problems were
excluded. OSCIR, T-ReCS, and Michelle were mounted on the Cassegrain
port of the telescopes so that north was up and east was left as
projected onto the detector.  In post-processing, images of the PSF
stars were de-rotated to match the telescope pupil PA when the galaxy
observations were observed.  Image rotation is necessary because the
projected angles of the telescope pupil (particularly the secondary
mirror supports) rotate during observations (or pointings), having a
significant effect on the low-level profile.

Both flux and point source function (PSF) standard observations were
made for flux and image quality calibrations through the same filters
used for each galaxy observation.  Flux standards were observed using
standard sources suggested from the Gemini web pages.  We find that,
consistent with other mid-IR observations, the uncertainty in the flux
determination is around 10\%.  PSF observations were made using the
same filters as the galaxy observations, immediately prior to or after
the galaxy observations and used an identical setup to accurately
sample the image quality.  Short PSF or flux standard observations are
comparable to longer source observations as the closed-loop active
optics of Gemini and the mechanical stability offered by the 4m at the
CTIO provides a similar PSF when taken at a similar telescope pointing
and time. Observations of PSF and flux standards through the night
showed a stable and consistent PSF.

We made three different photometric measurements of each nucleus. 
First, assuming that the peak emission represents the central source,
we scaled the PSF to this peak.  The scaled flux
integrated over a region that fully encompasses
the Airy disk is the ``diffraction-limited'' measurement,
where diameters of 2\arcsec{} are typical of the observations from
the 8.1m Gemini telescopes and 4\arcsec{} is appropriate for the observations
from the 4m Blanco telescope.
Second, we used a fixed
{\em physical} scale defined by the resolution of the most distant
galaxy, of 100 pc.  Third, we estimated the flux of the unresolved
nuclear component, the ``PSF-fitting'' measurement.  In this case, we
subtracted a scaled PSF from the galaxy image.  The PSF scaled to the
peak of the galaxy emission (the ``diffraction-limited'' measurement) 
represents the maximum contribution of the
unresolved source.  The residual of the total emission minus the
scaled PSF represents the host galaxy contribution. [A detailed study
of the extended near nuclear structures of the galaxies in our sample
will be the subject of a forthcoming paper (J. T. Radomski et al., in
preparation).]  Subtraction of a PSF scaled to match the peak of the
galaxy emission exactly results in an unrealistic minimum in the
residual host galaxy emission.  Instead, we reduce the scale of the
PSF to produce a ``flat'' profile in the residual to obtain the
unresolved fluxes reported in Table \ref{tab:lum}.  (See
\citealt{Ram09} for examples of this technique.)  We note that a flat
nuclear profile may not account for any residual stellar cusp at the
nucleus, but this effect will be small.  For reproducibility and
simplicity, we apply this approach to all of our sources.  We estimate
this method introduces a further uncertainty of 10--15\% in the final
flux measurements, which we add in quadrature to the flux calibration
errors.  In the case of the fixed spatial scale, a simple aperture was
placed over the galaxy and the flux extracted.  In all cases, a sky
annulus was used to subtract low-level, residual sky emission not
cancelled by the chopping and nodding.

While the observations were made through different filters in the
10\um{} window, we assume a flat spectral energy distribution through
the limited wavelength range they cover.  Thus, we make no corrections
to the measured fluxes in the comparisons below.  We expect this
simplified approach minimizes the introduction of additional
uncertainties, given that we do not know in detail the high spatial
resolution spectral energy distribution of each AGN.  We have few of
our own observations of Seyfert 1 galaxies, so we considered
supplementing these with results from the literature.  However, in
order to compare the different MIR measurement techniques, we quote
the statistical results from our sample alone.

\subsection{X-ray Measurements\label{subsec:xrdata}}
We take care that the X-ray measurements also isolate the AGN,
favoring high angular resolution data.  We prefer observations
close in time to the MIR observations, to avoid possible
complications of intrinsic AGN variability.  All the X-ray
fluxes are corrected for line of sight absorption (due to the
host galaxy, or Milky Way, or both) to determine intrinsic luminosity in
the 2--10 keV band, $L_X$.  
The underlying AGN continuum of 
Compton thick galaxies, those obscured by $N_H > 10^{24} \mathrm{\, cm^{-2}}$,
does not emerge directly below 10 keV.  In these cases, we prefer higher-energy
observations to determine the AGN luminosity.  These are generally uncontaminated
by stellar processes, despite the poor angular resolution.

Table \ref{tab:lum} lists the sample and basic data.  The tabulated
uncertainties reflect the range of observed luminosities, and genuine
variability is significant in Seyfert 1s.  The span of measured
Seyfert 2 luminosities is usually small for each AGN, and we adopt a
nominal 20\% uncertainty to account for intrinsic variability if none
is observed directly.  When detection of a Compton thick AGN at $E>
10$ keV is absent, the tabulated range indicates the uncertainty in
the correction to recover the intrinsic luminosity.  We use
measurements from the literature where they are available.  We present
our own X-ray analysis of three galaxies below, and we give special
consideration to the exceptional case of NGC 1068.

NGC 1068 is extremely optically thick, and the direct AGN emission is
not detected even above 10 keV \citep{Matt97}.  In this case, the
range of $L_X$ indicates the uncertainty in determining the intrinsic
luminosity.  We use the results of \citet{Lev06}, adopting the
estimate of $L_X$ from the Fe K$\alpha$ line equivalent width (EW) and
luminosity, and considering the fitted reflection model a lower limit,
for $D=14.4$ Mpc.  We use the estimate of \citet{Iwa97}, who consider
the properties of the ionized scattering region, as an upper limit on
$L_X$.

NGC 1566 is included in the XMM slew survey and detected in both hard
and soft bands in one observation \citep{Sax08}.  Assuming $\Gamma =
1.7$ (to apply the standard conversion from the survey count rate to
flux) we have absorption-corrected 2--10 keV flux $F_X= 5.8 (\pm
2)\times 10^{-12} \mathrm{\,erg\, cm^{-2}\, s^{-1}}$.

We fit the ASCA spectrum of NGC 3081, using source and background
screened events files from the Tartarus database.  We simultaneously
fit the 3--10 keV data from all four detectors, allowing for a
constant offset due to calibration uncertainty.  We ignore the softest
energies because they do not directly reveal the intrinsic AGN output.
Instead, they show significant line emission, suggesting a strong
thermal or photoionized contribution.  We fit the AGN as an absorbed
power law, fixing photon index $\Gamma = 1.9$, which is typical of
Seyfert galaxies.  We find $N_H = 6.3 (\pm 0.4) \times 10^{23} \mathrm{\, cm^{-2}}$, 
with absorption-corrected 2--10 keV flux 
$F_X = 3.8 (\pm 0.2) \times 10^{-11} \mathrm{\,erg\, cm^{-2}\, s^{-1}}$
averaged over all detectors.

We fit the Chandra spectrum of the nucleus of NGC 5728.  The data were
reduced as described in \citet{Lev06}, and we used a $2\arcsec$
nuclear aperture.  The direct continuum emission is strong.  Fitting
the data at $E> 3$keV and fixing $\Gamma = 1.9$, we find $N_H = 7.1
(\pm 2) \times 10^{23} \mathrm{\, cm^{-2}}$, and $F_X = 9.3 (\pm 2)
\times 10^{-12} \mathrm{\,erg\, cm^{-2}\, s^{-1}}$.  The neutral Fe
K$\alpha$ line is also strong, having EW = 790 eV.

\section{Results and Discussion}

\subsection{MIR and Bolometric Luminosity Correlations\label{subsec:corr}}
Consistent with earlier similar work \citep{Hor06,Hor08,Gan09}, 
we find overall that the X-ray and MIR luminosities
are well-correlated.  Because the X-ray measurements represent the intrinsic AGN
luminosity, with minimal contamination from other sources in the host galaxies, we
conclude that the MIR emission is strongly correlated with the AGN bolometric luminosity.
This result holds for all three methods we employ to measure the MIR emission, and 
considering both intrinsic variability of the X-ray emission and using only errors
on individual X-ray measurements. 
One simple way to quantify the strong correlation is 
to measure the 
unweighted mean ($\mu$) and standard deviation ($\sigma_r$) 
of $\log L_{MIR}/L_X$. 
Using the diffraction-limited MIR luminosity we find 
$\mu = 0.12$ and $\sigma_r =  0.6$;
in the PSF-fitted case we find
$\mu = 0.10$ and $\sigma_r = 0.7$,
and with the fixed physical aperture we find $\mu = 0.37$ and $\sigma_r = 0.5$.
These values are comparable to the results of 
\citet{Gan09}, although we measure greater dispersion in the present smaller sample.

Despite the intrinsic high resolution of the diffraction-limited
measurements, the differences with the PSF-fitted fluxes indicates
that even on large telescopes, diffraction-limited MIR observations do
not truly isolate the immediate dusty surroundings of AGNs, which are
confined to smaller ($< 10$ pc) scales \citep{Mas06}.  Given that we
cannot measure the torus by itself, the problem of $L_{MIR}^{dl}$ is
that the physical scale it encompasses varies among the different
galaxies.  Thus, while it offers a consistent technique from the
observational viewpoint, the physical correspondence changes.  The use
of a fixed physical aperture directly compensates for this
disadvantage, while reducing the effective resolution (for all but one
galaxy).  The PSF fitting offers the best determination of the flux of
the unresolved MIR-emitting torus, despite the variation of physical
scale from galaxy to galaxy.  Thus, this result best describes the
relationship between intrinsic AGN luminosity and the reprocessed
emission of the torus.

The MIR luminosity of NGC 4945  changes greatly with the three
different measurement techniques.  This galaxy contains a strong
starburst.  Other sources of MIR emission in addition to the AGN are
therefore significant on small scales, and these dominate even
$L_{MIR}^{dl}$, unlike in other galaxies.  Also, because NGC 4945 is
nearby, the fixed 100-pc aperture measurement covers a much larger
area (that contains strong emission sources) than the
diffraction-limited (36 pc) region does.

The differences among the various MIR measurements are larger for
the Seyfert 2 galaxies than for the Seyfert 1s.  Among the latter, the
PSF-fitting scales are all 90 or 100\%, producing nearly the same
result as the diffraction-limited measurements.  The fixed physical
aperture measurements remain similar for the Seyfert 1s.  In contrast,
PSF fitting of some Seyfert 2s implies that only a small fraction of
even the unresolved emission is the  torus.  In these galaxies,
other contributions remain significant on the 100 pc scale, as well.

\begin{figure}[htb]
\centerline{\includegraphics[width=\columnwidth]{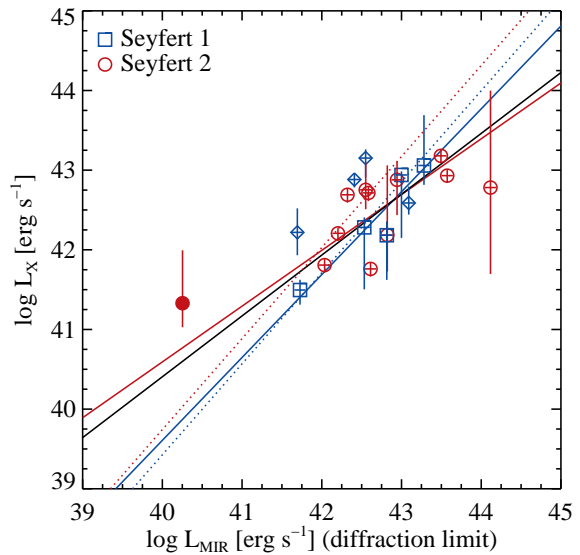}}
\caption{\label{fig:dl}
The MIR emission, which the reprocessing torus dominates, is strongly
correlated with AGN bolometric luminosity, which X-ray luminosity ($L_X$)
indicates.  Here we show diffraction-limited MIR measurements.  
Solid blue, red, and black lines show the linear fits to the  Seyfert 1 galaxies alone,
the Seyfert 2 galaxies alone, and all data combined, respectively.
The uncertainty in the type 1 relationship is large, so the resulting fit is not significantly
different from the type 2 or combined results. 
Observations of X-ray variability contribute to the effective error and are large for the
Seyfert 1s, while uncertainty of the intrinsic X-ray emission from the 
very absorbed AGNs accounts for the largest errors in the Seyfert 2s. Measurement 
uncertainty dominates the MIR errors. 
Additional data from the literature are plotted (blue diamonds); they do not contribute to the 
fits but are consistent with them.
The filled symbol is NGC 4945 (\S\ref{subsec:corr}), and 
dotted lines show theoretical predictions (\S\ref{subsec:models}).
}
\end{figure}

\begin{figure}[htb]
\centerline{\includegraphics[width=\columnwidth]{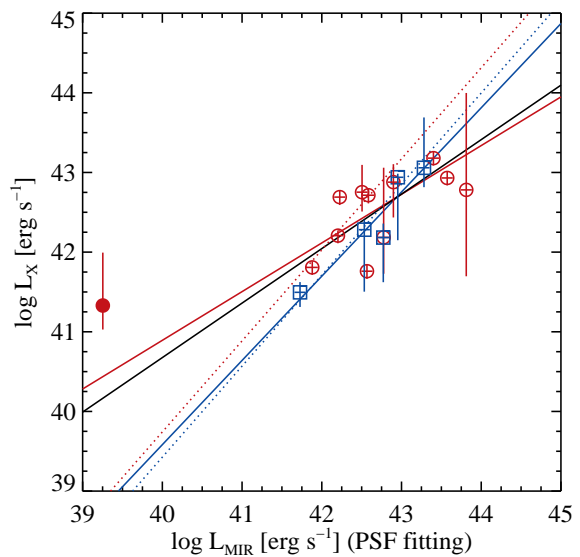}}
\caption{\label{fig:psf}
As in Figure \ref{fig:dl}, using PSF fitting of the MIR flux.
The resulting fits are similar, and the correlation is again significant.}
\end{figure}

\begin{figure}[tbh]
\centerline{\includegraphics[width=\columnwidth]{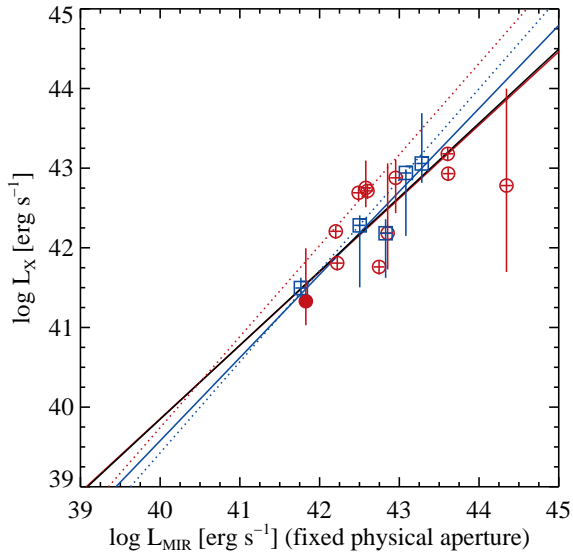}}
\caption{\label{fig:fixed}
As in Figure \ref{fig:dl}, where the MIR flux is measured within a fixed 100 pc aperture.
The subsequent fits are again similar to the other results, and the correlation is
again significant.
}
\end{figure}

Table \ref{tab:stats} lists the Spearman rank correlation coefficient
and the corresponding probability, considering Seyfert 1, Seyfert 2,
and total samples, and the three different MIR measurements
separately. The correlations of the Seyfert 2 and full samples are
significant; the less significant (though strong) correlation of the Seyfert 1 subsample alone
is a consequence of the small number of objects.  
Figures \ref{fig:dl} through \ref{fig:fixed} show the data and fitted
functions, where X-ray variability contributes to the measurement uncertainty.
The linear fits to the logarithmic luminosities are plotted separately
for the Seyfert 1 galaxies, Seyfert 2 galaxies, and the combined sample, 
where we use the bisector of
the fits of $L_{MIR}$ vs. $L_X$ and $L_X$ vs. $L_{MIR}$.  The
coefficients of the fits are listed in Table \ref{tab:stats}.  The
linear fits to the Seyfert 1 and Seyfert 2 galaxies are consistent
with each other, indicating that the MIR emission is essentially independent of
viewing angle and line of sight obscuration.  The uncertainty of the fits to the
limited Seyfert 1 sample alone is large, effectively a factor of 10 in the 
$L_{MIR}/L_X$ ratio.

Considering the combined Seyfert 1 and 2 samples, both the
diffraction-limited and PSF-fitting relationships suggest a dependence
on luminosity, in the sense that $L_X$ is reduced with greater MIR
emission.  We attribute this result to contamination in the MIR, with
additional components of dust emission associated with star formation
present even on these small scales.  Sources in addition to the AGN do
not affect the X-ray data we use, while they do contribute to the MIR
flux.  The galaxies having excess MIR emission (the ``contaminated''
nuclei) are preferentially those with higher MIR luminosity.
Moreover, this effect is reduced when using the fixed physical
aperture MIR measurement, in which the absolute amount of
contamination does not vary significantly among the different
galaxies.  We can interpret this result to mean that on scales of 100
pc, the star formation in all these galaxies is comparable. 
The luminosity dependence in the diffraction-limited and PSF-fitting
results implies that the {\em nuclear} star formation (arising on even
smaller scales) does vary.  
Star formation on such small scales in active galaxies 
has been directly measured at near-IR wavelengths, showing a range of
specific star formation rates 
\citep{Dav07}.
Considering the MIR/X-ray relationships,
while diffraction-limited or (especially) PSF-fitting measurements
best reveal the reprocessed AGN luminosity of an individual galaxy,
systematic differences introduce a luminosity dependence in the
MIR/X-ray correlation that we do not attribute to properties of the
AGN.

We note that the sense of the result of \citet{Gan09} is
similar to ours,
though weaker, 
with  $ L_X \sim  L_{MIR}^{0.90\pm0.06} $.
Thus, they do not claim any significant luminosity dependence in 
$L_{MIR}/L_X$.   While the luminosity range of their sample
is comparable, the average distance of their sources is greater.
The luminosity effects we find are strongest in the measurements that
best isolate the AGN torus, which
may not be apparent at lower effective spatial resolution.

In contrast,  \citet{Mai07} found a luminosity dependence in the
opposite sense, with MIR luminosity decreasing with increasing
intrinsic AGN luminosity.  That work investigated unobscured AGNs
only, using lower spatial resolution observations from the Spitzer Space Telescope.
Concentrating on more distant and higher luminosity AGNs, including
quasars, however, they attribute the dependence to genuine differences
in the torus as a function of luminosity.  Specifically, if
high-luminosity quasars partially clear their immediate surroundings,
having less dusty material available to reprocess the intrinsic
continuum would reduce the relative emergent MIR output.  However,
over the lower luminosity range we probe and utilizing high spatial
resolution, contamination by star-heated dust is a stronger effect.

If we extend the sample of Seyfert 1s to include diffraction-limited
measurements of nearby galaxies from the literature
\citep{Hor06,Hor08}, the fitted functions remain consistent with each
other over the different measurement techniques because of their large
uncertainties.  However, the strength and significance of the
diffraction limited Seyfert 1 correlation is reduced, with $\rho =
0.53$ and $P = 0.13$.  We plot these data in Figure \ref{fig:dl}, to
show their overall agreement with our results, but the fitted
relationships and statistical correlations of Table \ref{tab:stats} do
not use these measurements.

\begin{figure}[bht]
\includegraphics[width=\columnwidth]{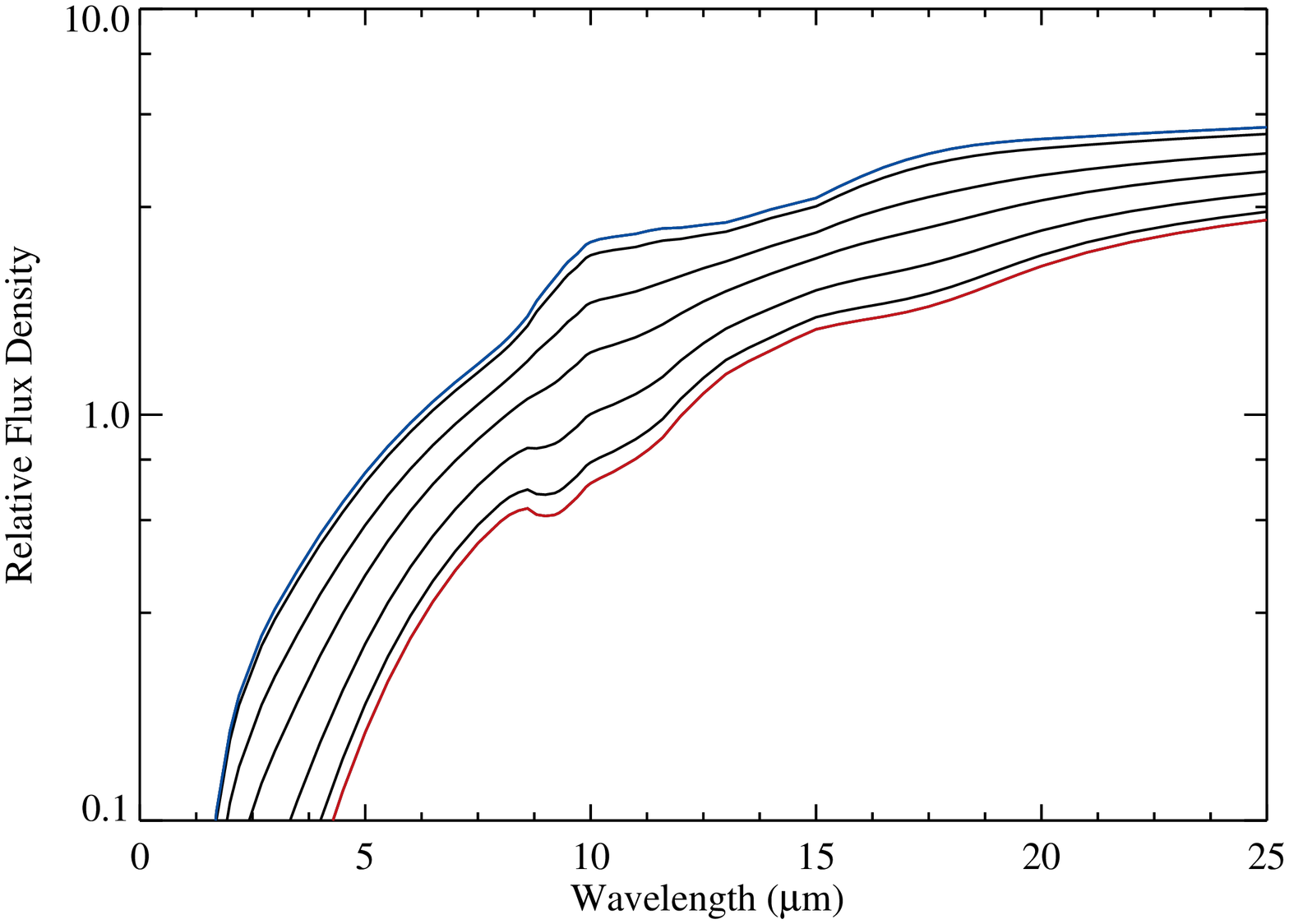}
\includegraphics[width=\columnwidth]{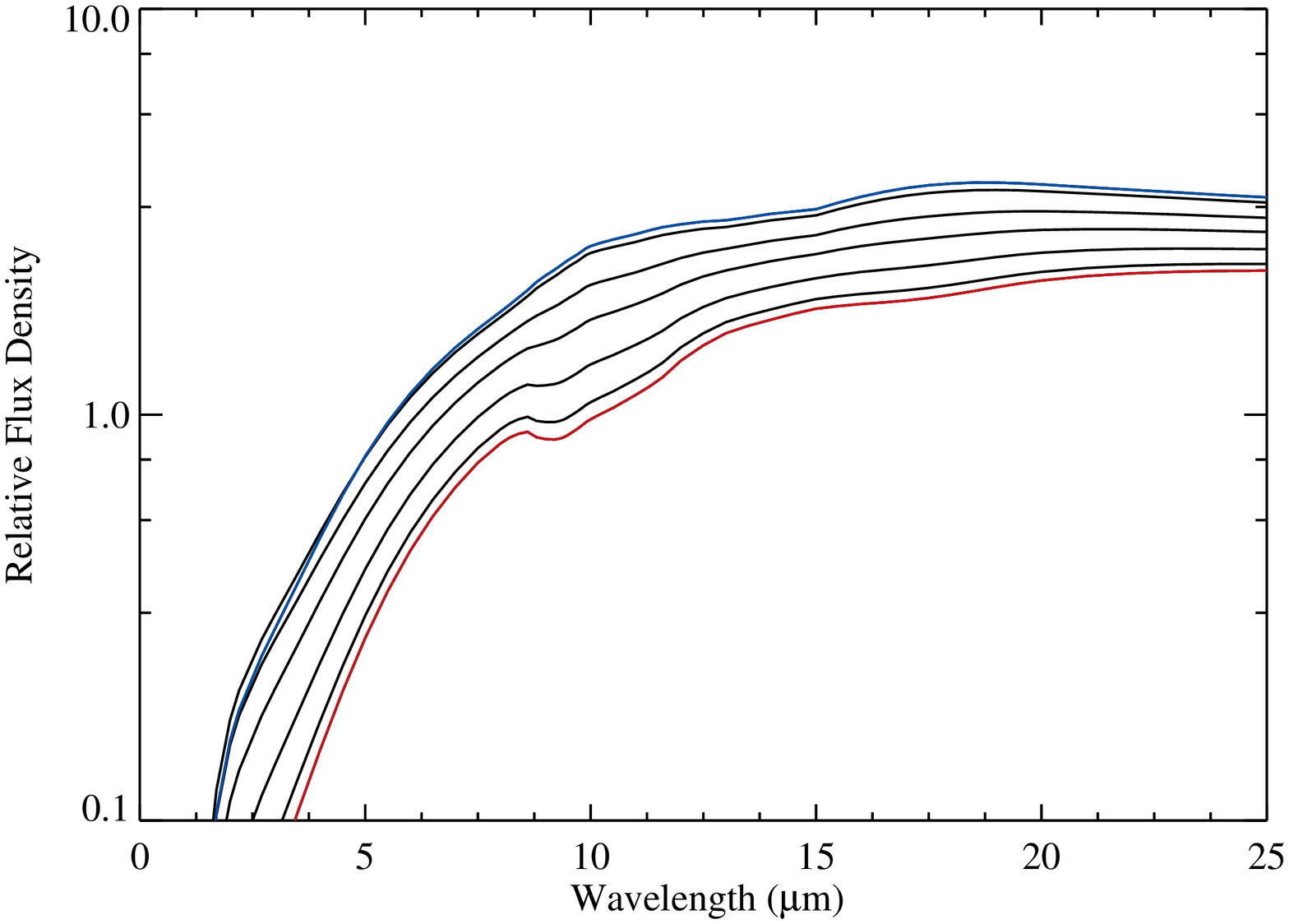}
\caption{\label{fig:specq12}
Clumpy torus model spectra \citep{Nen08b} from different viewing angles
show a range of silicate feature behavior (around 10\um)
while remaining roughly isotropic through the MIR continuum.
The inclination angles plotted are 0\degr{} (blue) through 30, 50, 60, 70, 80\degr{} to 
90\degr (red).
The spectra show very little variation from all type 1 (low inclination) views.
The model parameters are $\sigma=30\degr$, $N_0 = 4$, $\tau_V = 60$, and $Y=30$,
with  $q=1$ on the top, and $q=2$ on the bottom.
}
\end{figure}

\subsection{Isotropic MIR Emission from Clumpy Models\label{subsec:models}}

While a smooth optically and geometrically thick torus preserves the
essence of AGN unification, accounting for viewing-angle dependent
differences in the detectability of spectrally broad optical emission
lines, it cannot accommodate the observed MIR isotropy.  Detailed
radiative transfer models \citep[][]{Pier92,Gra94,Efs95} all indicate
significant differences in emergent MIR luminosity as a function of
viewing angle for a given AGN luminosity.  For any particular dust
configuration, changes in viewing angle result in MIR luminosity
differences of multiple orders of magnitude.  Considering a range of
simulated dust distributions, the resulting variation is even larger.
Fundamentally, the type 1 AGNs are preferentially brighter in the MIR
because they offer unobscured views of the hot optically thin surface
of the inner throat of the torus.  In contrast, obscuration of the
central engines of type 2 AGNs simultaneously hides this hottest (most
emissive) dust behind a large optical depth of cool material.

Clumpy torus models, in which the volume filling factor of the dusty
material remains small, can readily account for the observed MIR
isotropy while retaining the central features of AGN unification
\citep[e.g.,][]{Hon06,Nen08b,Scha08}.  The principal difference is
that these configurations can provide direct views of some hot and
some cold cloud faces from all lines of sight, even when the active
nucleus itself is hidden.  Thus, they reduce the sensitivity of MIR
emission to viewing orientation, while simultaneously allowing for
small tori and variability in line of sight obscuration.

We use the formalism and results of \citet{Nen08a,Nen08b} to interpret
these observations quantitatively.  Individual clouds are optically
thick in the $V$ band, given by $\tau_V$, with the cool silicate dust
of \citet{Oss92}.  The average number of clouds along an equatorial
ray is $N_0$, and they are distributed according to a power law in
radius, $\propto R^{-q}$, from the dust sublimation radius, $R_d$, to
an outer radius $R_o = Y R_d$.  The ``torus'' is concentrated in the
equatorial plane and has a smooth edge, with $\sigma$ setting the
distribution in elevation; the average number of clouds along the line
of sight $N_{LOS} = N_0 \exp(-(90\degr-i)^2/\sigma^2)$ at inclination
angle $i$ (measured from the pole).

Figure \ref{fig:specq12} presents the resulting MIR spectral energy
distributions as a function of inclination angle for some
characteristic parameter values.  They show extremely little variation
over all type 1 views ($i = 0 $ to 30\degr{}).  More importantly in
the context of the present observations, the MIR continuum flux
changes by only factors of a few from the extreme pole-on ($i =
0\degr$) to equatorial ($i = 90\degr$) lines of sight.  One further
consequence is that the silicate feature around 10\um{} can appear in
emission (especially from type 1 views) or absorption (especially from
higher inclinations), but it is generally weak, which is consistent
with spectral observations \citep{Hao07,Tho09}.

The emergent reprocessed MIR emission depends on the intrinsic AGN
luminosity.  We plot the theoretical MIR/X-ray correlations as dotted
lines in Figures \ref{fig:dl} through \ref{fig:fixed} for the
characteristic parameters of Figure \ref{fig:specq12} (adopting
$q=2$).  These model relationships have no luminosity dependence
(i.e., the slope of the lines is one), and show the offset factor of a
few between the extreme pole-on (blue) and equatorial (red) viewing
angles.  The absolute scale depends on the bolometric correction to
the X-ray luminosity.  The plotted lines assume that the
absorption-corrected 2--10 keV emission represents 5\% of the
intrinsic luminosity.  The theoretical predictions are extremely close
to the results of fitting the type 1 data alone for all MIR
measurements.  In contrast, most of the type 2 data lie to the right
of the model lines, which further supports the interpretation that
non-AGN MIR emission can be important even over very small angular
scales.

While the MIR emission is nearly isotropic, \citet{Nen08b} discuss
several ways the clumpy model parameters affect the sensitivity of MIR
emission to viewing angle.  In general, the most compact tori produce
the most isotropic emission.  The compactness may be achieved by
either reducing the outer torus extent (lowering the $Y$ parameter),
or by increasing the steepness of the radial power law, $q$, which
concentrates the clouds toward the inner regions.  Larger torus scale
heights provide more isotropic clump distributions, so increasing
$\sigma$ also reduces the sensitivity of MIR luminosity to viewing
angle.  Because increasing the number of clouds along radial rays
(increasing $N_0$) approaches the smooth models, these distributions
show greater MIR variation.  The combination of $N_0$, $\sigma$, and
$i$ determines the number of clouds along the line of sight, with
fewer clouds favoring greater variability and more views of
directly-illuminated cloud surfaces.

\begin{figure}[htb]
\centerline{\includegraphics[width=0.45\textwidth]{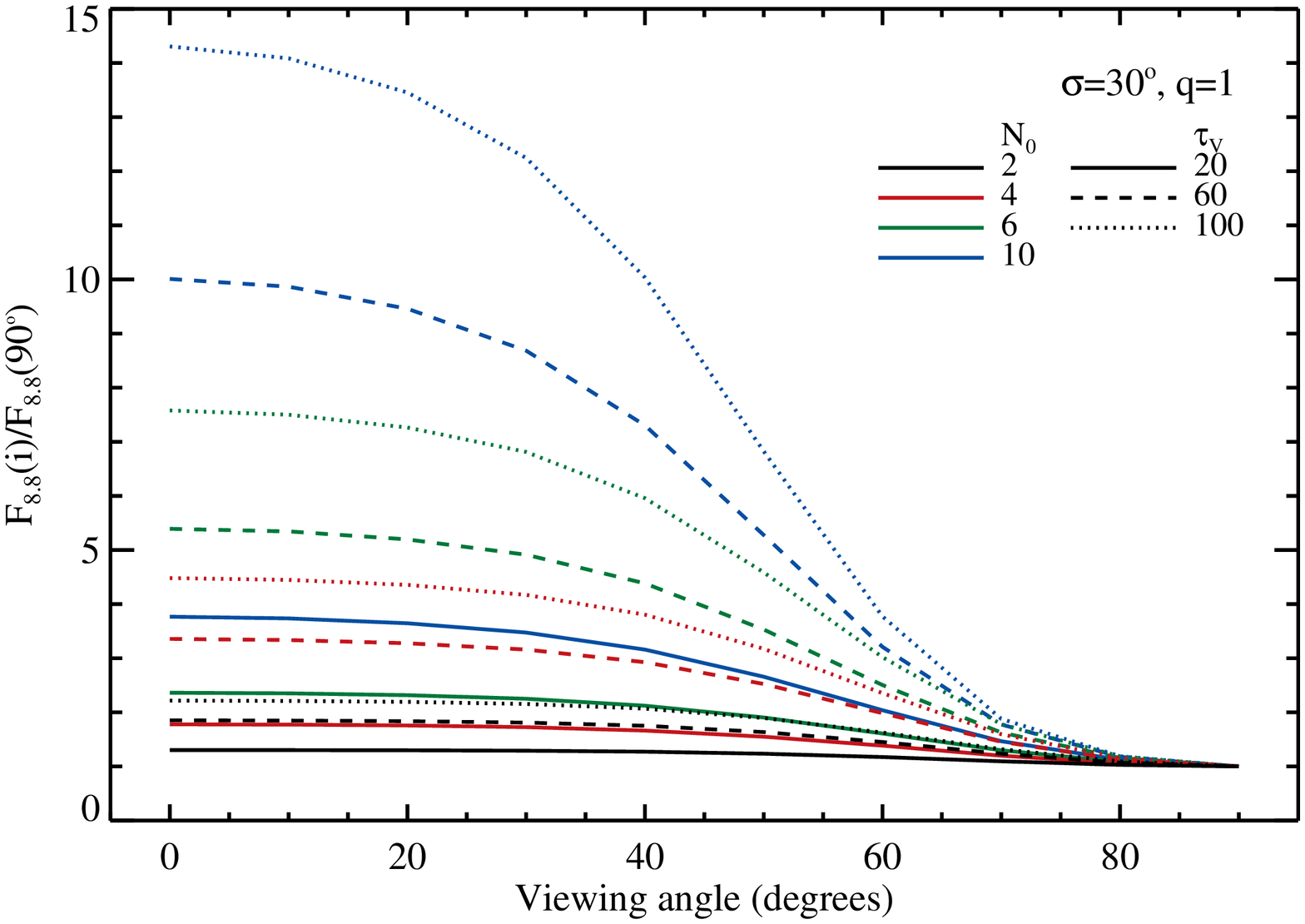}
\includegraphics[width=0.45\textwidth]{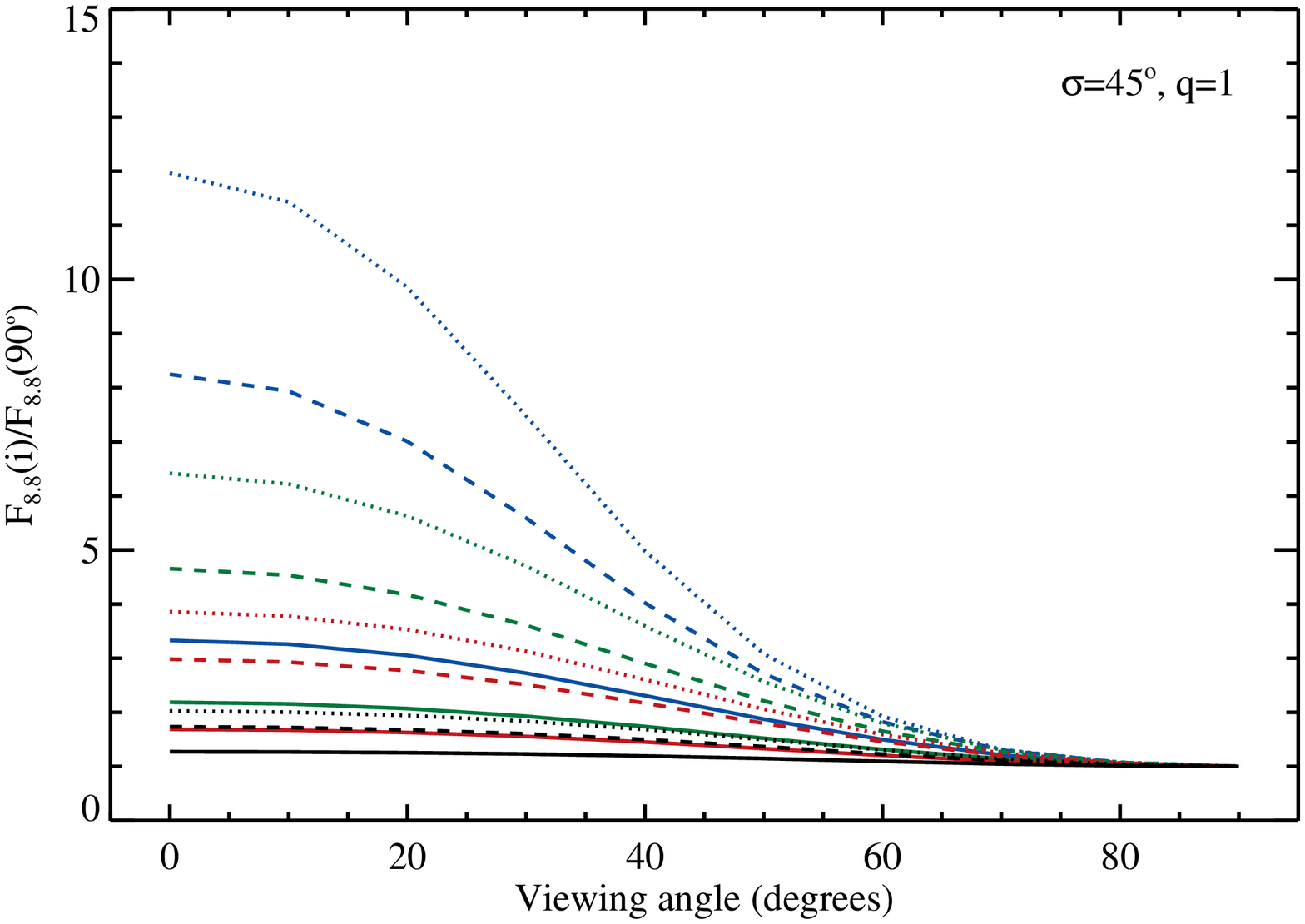}
}
\centerline{\includegraphics[width=0.45\textwidth]{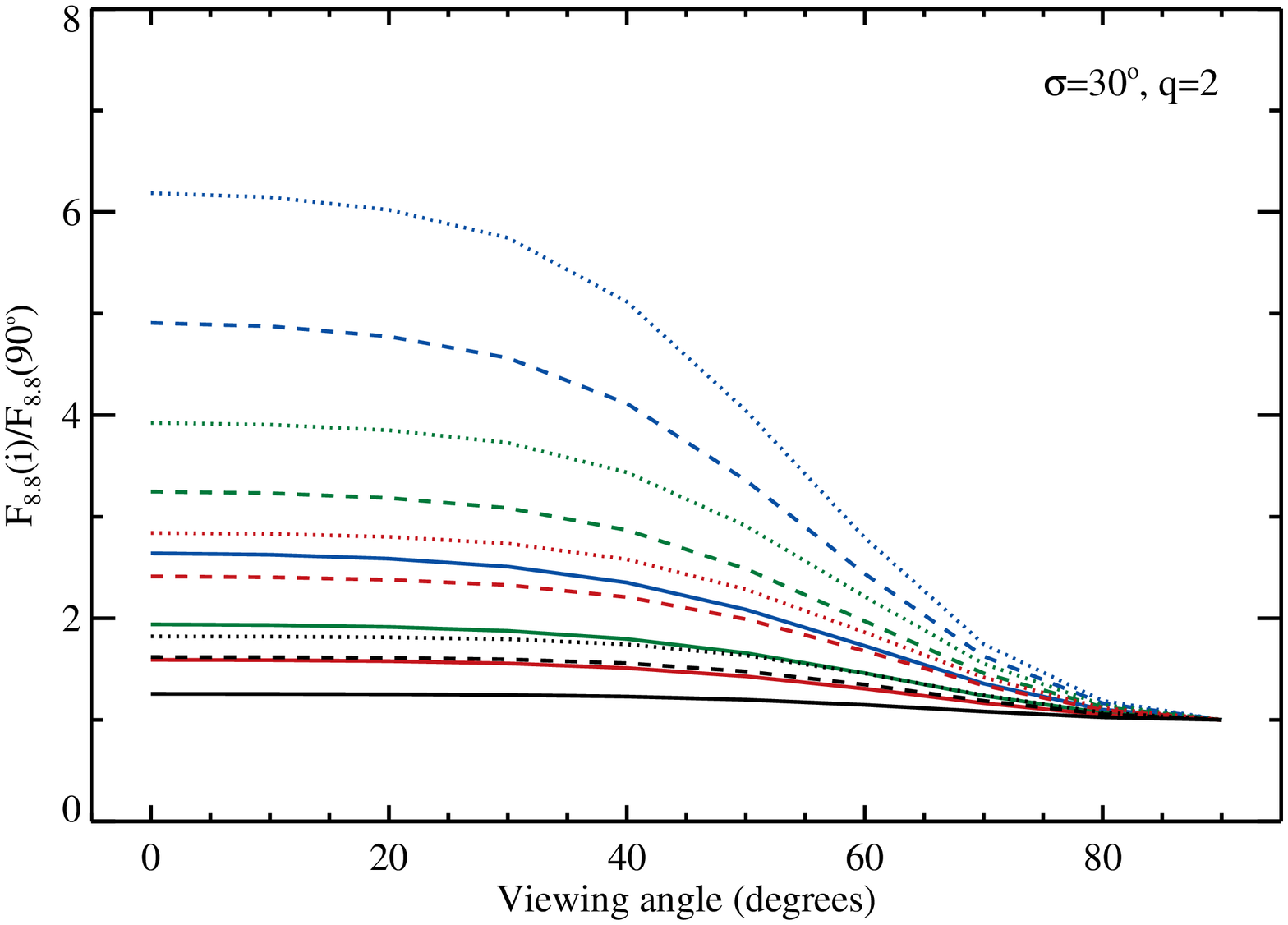}
\includegraphics[width=0.45\textwidth]{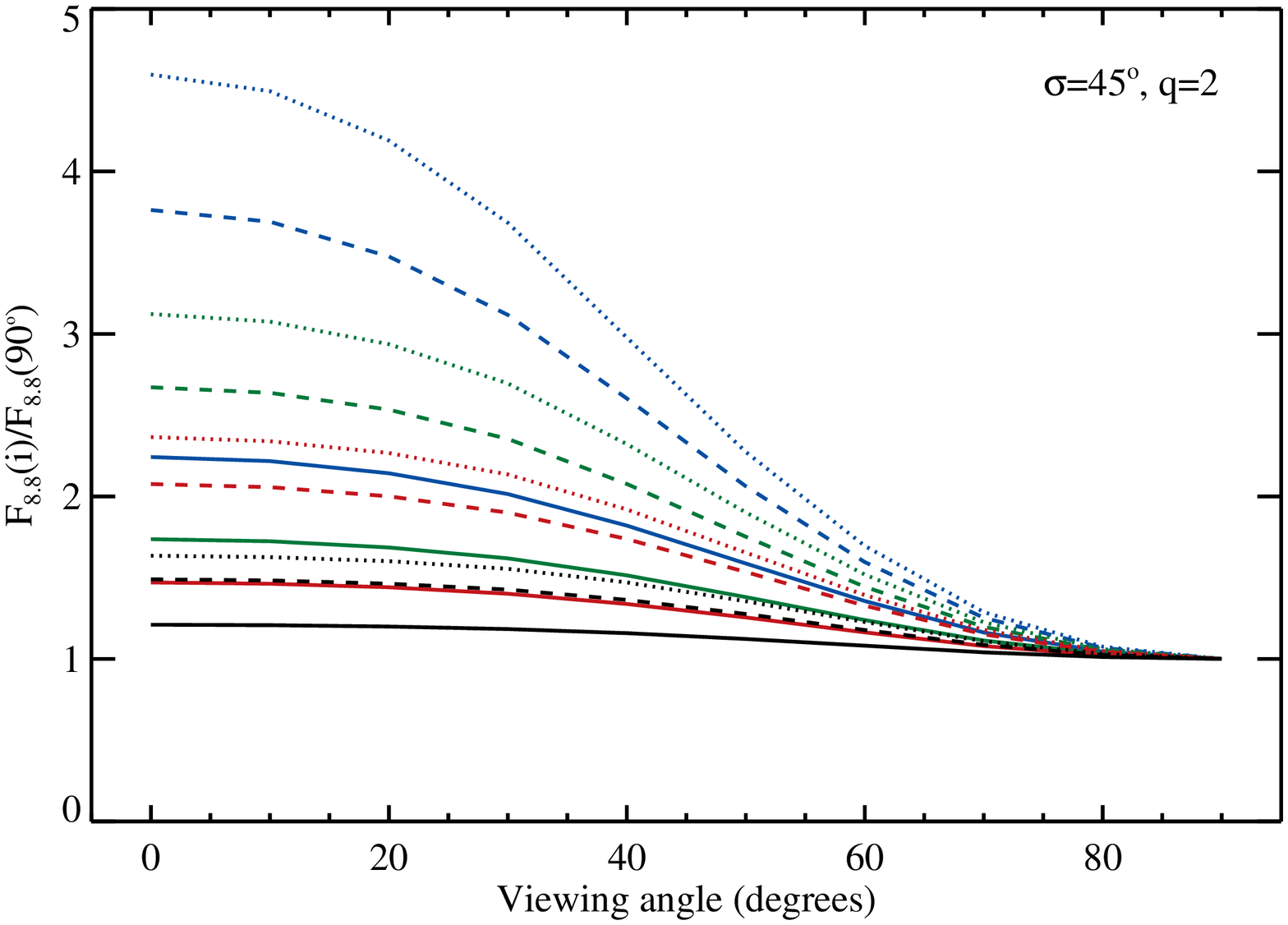}
}
\caption{\label{fig:iso}
  MIR flux variation as a function of viewing angle.  Each curve
  covers one clumpy dust distribution, and the flux is scaled to the
  value observed from the equatorial (edge-on) view.  
These results illustrate that the
  emergent 8.8\um{} flux does not vary by more than factors of about 5
  over all viewing angles, for typical parameter values ($N_0 \le 6$,
  $\tau_V \le 60$).  Making the torus effectively more compact
  (increasing $q$) or making the cloud distribution more spherically
  symmetric (increasing $\sigma$) increase the MIR isotropy, as do
  decreasing $N_0$ and $\tau_V$.  }
\end{figure}

Here we emphasize the uniformity of the emergent MIR flux.  We
consider a range of parameter values, with $N_0 = 1$--15, $\sigma =
15$--60, $q= 0$--3, and $\tau_V = 10$--200 in over 11,000 models.  The
absolute 8.8\um{} flux, $F_{8.8}$, does not vary by more than a factor
of 600 considering {\em all} these models, whereas smooth torus
calculations typically show even larger variation within a single
model realization.  For {\em any} given clumpy configuration,
$F_{8.8}$ does not change by more than a factor of 35 as a function of
viewing angle, and in more than 99\% of the cases, the variation is
less than a factor of 20.  For typical values of the model parameters,
changes in viewing angle do not result in MIR flux differences greater
than a factor of 5.  Figure \ref{fig:iso} illustrates these results,
showing the flux at 8.8\um{} as a function of viewing angle, scaled to
the 8.8\um{} flux from the equatorial (edge-on) view of the cloud distribution. 
Different panels show various values of $\sigma$ and $q$, and colors
and line styles indicate $N_0$ and $\tau_V$, respectively.  As
expected, isotropy at 8.8\um{} increases with increasing $\sigma$,
increasing $q$, decreasing $N_0$, and decreasing $\tau_V$, consistent
with the analysis of \citet{Nen08b}.

The efficiency of reprocessing intrinsic AGN emission to the MIR
is sensitive to the total number of clouds in the torus, which
$N_0$ and $\sigma$ govern.  These parameters also determine the
likelihood that the central engine is hidden.
The less efficient reprocessors are more likely to be classified
as type 1 AGNs, offering more unobscured lines of sight.
The effect of this selection bias is to reduce the net differences
in MIR emission between AGN types.  While for a given cloud distribution,
type 1 views show increased MIR flux over their type 2 counterparts,
the distributions that are more likely to be observed as type 1
have reduced MIR flux relative to the cloud configurations that are
more likely to be observed as type 2.  Observational samples remain
small, so conclusions are not statistically robust, but they do suggest 
differences in the cloud distributions between AGN types \citep{Ram09}.

Observations indicate that typical values for parameters
range over $N_0 \sim 5$--10, $\sigma \sim 30$--$45\degr$, and $\tau_V \sim 30$--100
\citep[and references therein]{Nen08b}.
Specifically considering the 10 and 18\um{} silicate features together, \citet{Tho09}
find 
$N_0 \le 6$, $\sigma=45\degr$, and   $\tau_V \sim 30$--60. 
The MIR-emitting tori of the clumpy models are small.  In terms of bolometric
AGN luminosity in units of $10^{45} \mathrm{\, erg\, s^{-1}}$, $L_{45}$,
the outer radius $R_o \approx 0.4 Y \sqrt{L_{45}}$ pc, with $Y = 30$ or less.
Even for the most luminous Seyfert galaxy of the sample, NGC 3281,
$R_o \le 7$ pc,
assuming $L_{bol} = 20 L_X$ \citep{Elv94}.
The torus is not resolved in direct imaging, despite
 the good angular resolution of  large ground-based telescopes.
NGC 4151 offers the most favorable combination of luminosity and proximity,
yet here $R_o$ corresponds to an angular scale $\le 0.07\arcsec$.
This value is well below the diffraction limit of $0.27\arcsec$ at $8.8\um$ on an 8.1 m telescope.
Intriguingly, resolving the torus may be possible on a 30m telescope, where
$R_{8.8} = 0.074\arcsec$. 
MIR interferometry with MIDI at the VLTI 
currently offers sufficient resolution and has been
used successfully to isolate AGN tori \citep{Jaf04,Tri07,Rab09}.  
However, this technique requires
very bright targets, so the number of AGNs available for study with
this facility is limited, with no more than a dozen being
feasible in total \citep{Rab08}.

\citet{Hor08} and \citet{Gan09} attempt to address possible
contamination in their comparable measurements by considering their
ability to resolve emission over larger scales.  Specifically, they
define sources to be ``well-resolved'' if the physical scale of an
observation's angular resolution is less than $560 R_d$, which is many
times larger than the outer torus radius $R_o$.  The clumpy model
predictions and basic considerations of the temperature of dust that
emits significantly in the MIR argue instead that the relevant size
scales are much smaller, so ``resolution'' on these scales of $560
R_d$ does not limit the measurements to the torus alone.  However, the
empirical support these authors find for the utility of this
resolution limit may be a consequence of dust heated by stars and star
formation, which can dominate on scales larger than their 100 pc
resolution limit ($\approx 560 R_d$ for their typical 
$L_X=10^{43} \mathrm{\, erg\, s^{-1}}$).  We find these non-AGN contributions to be
significant on all scales.  The flux in the fixed physical aperture is
(on median) 20\% higher than the PSF-fitted value.  On scales smaller
than 100 pc, the nuclear star formation produces the observed
luminosity dependence in the MIR/X-ray relationship.

One completely different explanation for the observed isotropy is
that the torus is simply optically thin in the MIR, as 
\citet{Hor06} considered.
However, an optically thin torus would show 10\um{} silicate emission
in all cases, contrary to  spectral observations.
While silicate absorption attributable to the torus is generally weak,
it is typical of type 2 AGNs, which requires optically thick material
along the line of sight.  (A smaller column in a cold foreground screen
could alternatively account for the absorption, but this could not
simultaneously be the source of the IR emission.)
Furthermore, the 
Compton thick AGNs require total $\tau_V > 800$ along the line of sight
(assuming standard gas-to-dust ratios).  This value corresponds to 
total $\tau > 14$ at 8.8um, although dust-free gas (located within the
dust sublimation radius) may also contribute to their X-ray absorption. 

\section{Conclusions}

Intrinsic X-ray and reprocessed MIR luminosity of the active nuclei of
Seyfert galaxies are strongly correlated.  Emphasizing measurements at
high spatial resolution, we find a luminosity dependence on the
correlation over the range we probe when using two MIR measurement
techniques that help to isolate the central unresolved emission.  We
attribute this effect to the presence of additional sources in the
MIR, such as by emission by dust that young stars heat.  Because this
result emerges only using the two smallest scale measurements, it
implies that significant nuclear star formation is present even on
very small (sub-100 pc) spatial scales in these AGNs.  However, we
find no luminosity dependence in the MIR fluxes measured in fixed 100
pc apertures, suggesting that the stellar contributions on these
scales are comparable in all galaxies.

We find no significant difference in the correlations between Seyfert
1 and Seyfert 2 galaxies, given the large ($\sim$ factor of 10)
uncertainty in the Seyfert 1 $L_{MIR}/L_X$ ratio.  Thus, we conclude
that the MIR emission of the AGN torus is nearly isotropic on all the
small scales we probe here.  We interpret these results as support for
clumpy torus models, in which the MIR emission is insensitive to
viewing angle because both illuminated and dark cloud faces contribute
in part to observed flux over all viewing angles.  In contrast, smooth
obscuration models produce significantly stronger (by several orders
of magnitude) MIR emission from type 1 AGNs, which exclusively allow
direct views of the hottest dust surfaces.

We use numerical radiative transfer calculations to demonstrate
explicitly the small differences in the spectral energy distributions
for extreme pole-on (type 1) and edge-on (type 2) views of a clumpy
torus.  For typical model parameters, the MIR flux changes by less
than a factor of 5 over all viewing angles.  Considering a wide range
of {\em different} cloud distributions, the observed variation is
nearly always less than a factor of 20.  Several factors that
contribute to increasing the resulting isotropy include increasing the
spherical symmetry of the cloud distribution, reducing the number of
clouds along radial rays, and making the clumpy torus more compact.
Furthermore, flux differences as a function of viewing angle decline
at longer wavelengths.

While the clumpy models predict nearly isotropic MIR emission,
selection effects may further contribute to minimizing the observed
difference between AGN types, whose classification is sensitive to
viewing angle.  The tori of the AGNs that are most likely to be
observed without obscuration reprocess the intrinsic emission less
efficiently than those that contain more dusty clouds (which are more
likely to be classified as type 2s).  Consequently, the relative MIR
flux of observed type 1s may be preferentially suppressed, while the
observed type 2s may be preferentially MIR-enhanced.

An optically thin torus could account for the MIR isotropy, but it could
not produce the observed spectral differences.  Specifically, silicate absorption
is typical of Seyfert 2s, and weak silicate features (in emission or absorption)
are characteristic of Seyfert 1s.
The clumpy torus is not optically thin in the MIR, which
allows for these spectroscopic differences.  Because both directly-illuminated
and dark cloud faces contribute to the net observed spectrum, the resulting
silicate features are always weak.

\acknowledgments
NAL thanks the University of Florida Department of Astronomy,
the Instituto de Astrof\'isica de Canarias, and the University of Oxford Department of Physics
for their hospitality during this work. 
Based on observations obtained at the Gemini Observatory, which is operated by the
Association of Universities for Research in Astronomy, Inc., under a cooperative agreement
with the NSF on behalf of the Gemini partnership: the National Science Foundation (United
States), the Science and Technology Facilities Council (United Kingdom), the
National Research Council (Canada), CONICYT (Chile), the Australian Research Council
(Australia), Minist\'erio da Ci{\^{e}}ncia e Tecnologia (Brazil) 
and Ministerio de Ciencia, Tecnolog\'ia e Innovaci\'on Productiva  (Argentina).
This work uses data from programs
GS-2005A-Q-6,  GS-2005B-DD-6, GN-2006A-Q-11, GN-2006A-Q-30, and GS-2006A-Q-62. 
We acknowledge the work of the University of Florida Department of Astronomy instrument group
for the OSCIR observations, especially that of Robert Pi\~na and Kevin Hanna. 
This research has made use of the NASA/IPAC Extragalactic Database
(NED) which is operated by the Jet Propulsion Laboratory, California
Institute of Technology, under contract with the National Aeronautics
and Space Administration. 
This research has also made use of data obtained from the High Energy
Astrophysics Science Archive Research Center (HEASARC), provided by
NASA's Goddard Space Flight Center, including the Tartarus (Version
3.1) database, created by Paul O'Neill and Kirpal Nandra at Imperial
College London, and Jane Turner at NASA/GSFC. Tartarus is supported by
funding from PPARC, and NASA grants NAG5-7385 and NAG5-7067.
NAL  acknowledges work supported by the NSF 
under Grant No. 0237291, and
CP  acknowledges work supported by the NSF 
under Grant No. 0206617.

\begin{deluxetable}{lccccr}
\tabletypesize{\footnotesize}
\tablewidth{0pt}
\tablecaption{Observation Log\label{tab:obs}}
\tablehead{
\colhead{Galaxy}
&\colhead{Instrument}
&\colhead{Telescope}
&\colhead{Observation Date}
&\colhead{Filter}
&\colhead{On-Source Time (s)\tablenotemark{a}}
}
\startdata
Circinus & T-ReCS &   Gemini S & 2004 Feb 01 & Si2 & 565   \\
IC 5063  & T-ReCS &   Gemini S & 2005 Jul 18 & Si2 & 130   \\ 
NGC 1068\tablenotemark{b} & Michelle & Gemini N & 2004 Aug 10 & Si5 &  4   \\ 
&&&2005 Jan 31  & Si5 &  4   \\  
NGC 1365 & OSCIR  &       CTIO & 1998 Feb 12 & N   & 482   \\ 
NGC 1386 & T-ReCS &   Gemini S & 2003 Dec 06 & N   & 217   \\ 
NGC 1566 & T-ReCS &   Gemini S & 2005 Sep 18 & Si2 & 152    \\ 
NGC 2992 & Michelle & Gemini N & 2006 May 05 & N$^\prime$  & 730   \\ 
NGC 3081 & T-ReCS &   Gemini S & 2006 Jan 25 & Si2 & 130   \\ 
NGC 3227 & Michelle & Gemini N & 2006 Apr 07 & N$^\prime$  & 300  \\ 
NGC 3281 & T-ReCS &   Gemini S & 2004 Jan 30 & N   & 261   \\ 
NGC 4151 & OSCIR  &   Gemini N & 2001 May 07 & N   & 360  \\ 
NGC 4388 & Michelle & Gemini N & 2006 May 12 & N$^\prime$  & 549   \\ 
NGC 4945 & T-ReCS &   Gemini S & 2006 Mar 17 & Si2 & 261   \\ 
NGC 5506 & Michelle & Gemini N & 2006 Apr 06 & N$^\prime$  & 546   \\ 
NGC 5728 & T-ReCS &   Gemini S & 2005 Jul 08 & Si2 & 130   \\ 
NGC 7172 & T-ReCS &   Gemini S & 2004 May 13 & N   & 305 \\ 
NGC 7582 & OSCIR  &   Gemini S & 2001 Dec 13 & N   & 608   \\ 
\enddata
\tablenotetext{a}{In guided beam only.}
\tablenotetext{b}{Both observations were used for flux calibration, 
 with spatial profile measurements of only the 2004 data, obtained under better seeing conditions.}
\end{deluxetable}

\begin{deluxetable}{llrllcccrc}
\tabletypesize{\scriptsize}
\tablewidth{0pt}
\tablecaption{Nearby Seyfert Galaxy Sample\label{tab:lum}}
\tablecolumns{10} 
\tablehead{
\colhead{Galaxy}&
\colhead{z}&
\colhead{D}&
\colhead{Type}&
\colhead{$L_X$\tablenotemark{a}}&
\colhead{$L_{MIR}^{dl}$\tablenotemark{b}}&
\colhead{$L_{MIR}^{psf}$\tablenotemark{c}}&
\colhead{$L_{MIR}^{phys}$\tablenotemark{d}}&
\colhead{PSF Scale}&
X-ray Reference(s)\\
&&\colhead{(Mpc)}&
&\colhead{log (erg s$^{-1}$)}
&\colhead{log (erg s$^{-1}$)}
&\colhead{log (erg s$^{-1}$)}
&\colhead{log (erg s$^{-1}$)}
&\colhead{(\%)}\\
}
\startdata
Circinus & 0.0014 &  4.0 &  2.0 &  $41.76^{+0.08}_{-0.10}$  & 42.61 &  42.57 & 42.75  & 90 & 1 \\
IC 5063  & 0.0113 & 47.9 &  2.0 &  $42.93^{+0.08}_{-0.09}$  & 43.57 &  43.57 & 43.61  & 100 & 2 \\
NGC 1068 & 0.0038 & 14.4 &  2.0 &  $42.78^{+1.22}_{-1.08}$  & 44.12 &  43.81 & 44.34  & 50 & 3, 4 \\
NGC 1365 & 0.0055 & 23.0 &  1.8 &  $42.18^{+0.18}_{-0.56}$  & 42.82 &  42.77 & 42.83  & 90 & 5 \\
NGC 1386 & 0.0029 & 12.2 &  2.0 &  $41.81^{+0.08}_{-0.09}$  & 42.04 &  41.88 & 42.22  & 70 & 3 \\
NGC 1566 & 0.0050 & 21.2 &  1.0 &  $41.50^{+0.13}_{-0.18}$  & 41.73 &  41.73 & 41.77  & 100 & 6 \\ 
NGC 2992 & 0.0077 & 32.6 &  2.0 &  $42.18^{+0.88}_{-0.46}$  & 42.82 &  42.78 & 42.85  & 90 & 7 \\
NGC 3081 & 0.0080 & 33.7 &  2.0 &  $42.71^{+0.08}_{-0.10}$  & 42.58 &  42.58 & 42.60  & 100 & 6 \\
NGC 3227 & 0.0039 & 16.3 &  1.5 &  $42.28^{+0.12}_{-0.78}$  & 42.53 &  42.53 & 42.50  & 100 & 8, 9 \\
NGC 3281 & 0.0107 & 45.1 &  2.0 &  $43.18^{+0.08}_{-0.09}$  & 43.49 &  43.40 & 43.61  & 80 & 10 \\
NGC 4151 & 0.0033 & 14.0 &  1.5 &  $42.94^{+0.04}_{-0.79}$  & 43.00 &  42.95 & 43.08  & 90 & 11, 12, 13 \\
NGC 4388 & 0.0084 & 35.5 &  2.0 &  $42.88^{+0.23}_{-0.44}$  & 42.94 &  42.90 & 42.95  & 90 & 14, 15\\
NGC 4945 & 0.0019 &  3.7 &  2.0 &  $41.33^{+0.66}_{-0.30}$  & 40.25 &  39.25 & 41.83  & 10 & 16, 17 \\
NGC 5506 & 0.0062 & 26.1 &  1.9 &  $43.06^{+0.63}_{-0.24}$  & 43.28 &  43.28 & 43.28  & 100 & 18, 9\\
NGC 5728 & 0.0094 & 39.5 &  2.0 &  $42.21^{+0.07}_{-0.10}$  & 42.20 &  42.20 & 42.20  & 100 & 6 \\
NGC 7172 & 0.0087 & 36.7 &  2.0 &  $42.75^{+0.34}_{-0.24}$  & 42.55 &  42.51 & 42.58  & 90 & 19, 20, 21\\
NGC 7582 & 0.0053 & 22.2 &  2.0 &  $42.69^{+0.08}_{-0.10}$  & 42.32 &  42.23 & 42.49  & 80 & 22 \\
\sidehead{Additional galaxies\tablenotemark{e}} 
MCG -6-30-15 & 0.0077 & 32.7 &  1.2 &  $42.59^{+0.21}_{-0.15}$  & 43.09  &\nodata&\nodata & \nodata & 23 \\
NGC 3783 & 0.0097 & 41.1 &  1.5 &  $43.15^{+0.11}_{-0.24}$  & 42.55  &\nodata&\nodata & \nodata & 24, 25 \\
NGC 4593 & 0.0090 & 38.0 &  1.0 &  $42.88^{+0.08}_{-0.10}$  & 42.41  &\nodata&\nodata & \nodata & 26 \\
NGC 7314 & 0.0050 & 21.1 &  1.9 &  $42.22^{+0.30}_{-0.29}$  & 41.69  &\nodata&\nodata & \nodata & 27 \\
\enddata
\tablenotetext{a}{Absorption-corrected 2--10 keV X-ray luminosity.}
\tablenotetext{b}{Diffraction-limited aperture measurement.}
\tablenotetext{c}{Measurement from PSF fitting.}
\tablenotetext{d}{Fixed physical (100 pc) aperture measurement.}
\tablenotetext{e}{From Horst et al. (2008).}
\tablerefs{
(1) Soldi et al. 2005; 
(2) Turner et al. 1997; 
(3) Levenson et al. 2006; 
(4) Iwasawa et al. 1997; 
(5) Risaliti et al. 2005;
(6) This work; 
(7) Yaqoob et al. 2007; 
(8) Lamer et al. 2003; 
(9) Uttley \& McHardy 2005; 
(10) Vignali \& Comastri 2002; 
(11) Beckmann et al. 2005; 
(12) Schurch \& Warwick 2002; 
(13) Yaqoob \& Warwick 1991; 
(14) Elvis et al. 2004; 
(15) Iwasawa et al. 2003; 
(16) Done et al. 2003; 
(17) Madejski et al. 2000;
(18) Lamer et al. 2000; 
(19) Awaki et al. 2006; 
(20) Akylas et al. 2001; 
(21) Smith \& Done 1996; 
(22) Turner et al. 2000; 
(23) Miniutti et al. 2007;
(24) De Rosa et al. 2002; 
(25) Kaspi et al. 2002; 
(26) Reynolds et al. 2004; 
(27) Yaqoob et al. 2006. 
}
\end{deluxetable}

\begin{deluxetable}{l*{3}{lccrc}}
\tablewidth{0pt}
\tablecaption{Correlation Results\label{tab:stats}}
\tablehead{
&\multicolumn{4}{c}{Diffraction Limited}&&
\multicolumn{4}{c}{PSF Fitting}&&
\multicolumn{4}{c}{Physical Aperture}\\
\cline{2-5}
\cline{7-10}
\cline{12-15}
\colhead{Sub-Sample} 
&\colhead{$\rho$}&\colhead{$P$}
&\colhead{$a$}
&\colhead{$b$}&
&\colhead{$\rho$}&\colhead{$P$}
&\colhead{$a$}
&\colhead{$b$}&
&\colhead{$\rho$}&\colhead{$P$}
&\colhead{$a$}
&\colhead{$b$} 
}
\startdata
All galaxies & 
 0.78 & 1.8e-03 & $ 0.76 \pm  0.06 $ &    9.9 && 
 0.79 & 1.5e-03 & $ 0.68 \pm  0.06 $ &   13.3 && 
 0.77 & 2.0e-03 & $ 0.93 \pm  0.06 $ &    2.7 \\ 
Seyfert 1 &  
 0.90 & 7.2e-02 & $ 1.0  \pm  0.2  $ &   -2.0 && 
 0.90 & 7.2e-02 & $ 1.1  \pm  0.2  $ &   -2.7 && 
 0.90 & 7.2e-02 & $ 1.0  \pm  0.2  $ &   -2.1 \\
Seyfert 2 &                         
 0.74 & 1.4e-02 & $ 0.70 \pm  0.07 $ &   12.6 && 
 0.78 & 1.0e-02 & $ 0.61 \pm  0.06 $ &   16.4 && 
 0.73 & 1.6e-02 & $ 0.92 \pm  0.07 $ &    3.0 \\ 
\enddata
\tablecomments{{S}pearman rank correlation coefficient, $\rho$,
  and two-sided probability, $P$.
 {L}inear fit parameters, for $\log (L_X) = a \log(L_{MIR})+b.$}
\end{deluxetable}

\clearpage

\end{document}